\documentclass[prb,twocolumn,superscriptaddress]{revtex4}

\usepackage{graphicx}
\usepackage{subfigure}
\usepackage{amsmath}
\usepackage{braket}

\usepackage{color}
\DeclareMathOperator{\sig}{\Sigma}
\DeclareMathOperator{\sigT}{\Tilde{\Sigma}}
\newcommand{\Eprod}[3]{\ensuremath{\sideset{^{#1}}{_{#3}^{#2}}\sig}}\newcommand{\EprodT}[3]{\ensuremath{\sideset{^{#1}}{_{#3}^{#2}}\sigT}}\newcommand{\note}[1]{\textcolor{red}{\bf{/*#1*/}}}
\renewcommand{\note}[1]{}

\newcommand{\simpleone}{\ensuremath{{1\!\!1}}}
\newcommand{\one}{\ensuremath{
   \mbox{\small
  $\displaystyle
  1
  $} 
\!\!1}}

\newcommand{\As}[2]{\ensuremath{A^{[#1]}_{#2}}}

\begin{document}
 \title{Efficient Matrix Product State Method for periodic boundary conditions} 
 \author{Peter Pippan}
 \affiliation{Institut f\"ur  Theoretische Physik, Technische Universit\"at Graz, A-8010 Graz, Austria}
 \author{Steven R. White}
 \affiliation{Department of Physics and Astronomy University of California, Irvine, CA 92697}
 \author{Hans Gerd Evertz}
 \affiliation{Institut f\"ur  Theoretische Physik, Technische Universit\"at
  Graz, A-8010 Graz, Austria}
 \date{\today}

\begin{abstract}
We introduce an efficient method to calculate the ground state of one-dimensional
lattice models with periodic boundary conditions. 
The method works in the representation of Matrix Product States (MPS), 
related to the Density Matrix Renormalization Group (DMRG) method.
It improves on a previous approach by Verstraete {\it et al.}. 
We introduce a factorization procedure for long products of MPS matrices, 
which reduces the computational effort from $m^5$ to $m^3$, where
$m$ is the matrix dimension,
and $m \simeq 100 - 1000$ in typical cases.
We test the method on the $S=\frac{1}{2}$ and $S=1$ Heisenberg chains.
It is also applicable to non-translationally invariant cases.
The new method makes ground state calculations with periodic boundary conditions
about as efficient as traditional DMRG calculations for systems with open boundaries.
\end{abstract}

\maketitle
One of the most severe problems in condensed matter theory is the exponential
growth of the Hilbert space with system size. This limits many methods such
as exact diagonalization. One strategy that overcomes these difficulties is to
approximate the ground state in some reduced Hilbert space. 

The \emph{Density Matrix Renormalization Group}
(DMRG)~\cite{white:DMRG_prl,white:DMRG_prb,schollwock:DMRG_review} is one prominent 
example of such methods.
By tracing out ''unimportant'' degrees of freedom, the real
ground state is approximated in a much smaller space. DMRG works much better 
for open boundary conditions (obc) than for
periodic boundary conditions (pbc). In the worst case where the correlation length
is much smaller than the system size, if the obc system 
needs $m_{\rm obc}$ states per block for a given accuracy,
the pbc system needs $O(m_{\rm obc}^2)$. Since the calculation time scales as $m^3$, the
comparable time for pbc is $O(m^6_{\rm obc})$.
However, systems with obc naturally suffer from edge effects like Friedel oscillations.
An efficient method for pbc would be highly desirable. For example, it would 
make finite size scaling easier, and allow the direct representation
of finite momentum eigenstates
\cite{porras:spectra_2005,ostlund:thermodynamic_limit_DMRG,rommer:mps}.

It can be shown that the ground state produced by DMRG can quite naturally be
written in terms of a so called \emph{matrix product state}
(MPS)~\cite{ostlund:thermodynamic_limit_DMRG,rommer:mps}
for both obc and pbc.
The original work presented an inefficient method for computing the MPS,
which could not compete with DMRG.
Recently, a number of new algorithms utilizing the MPS state directly have been
introduced which are efficient and greatly extend the reach of DMRG/MPS techniques
\cite{vidal_prl_03,verstraete:DMRG_and_pbc,porras:spectra_2005,
paredes:random_mps,verstraete:2d_mps,vidal_translational_invariant,
QMC_MPS1,QMC_MPS2},
including the simulation of random systems or a generalization to 2D--systems.
In the present paper we investigate an algorithm presented
in Ref.~\onlinecite{verstraete:DMRG_and_pbc}  for an MPS treatment of pbc systems.
Within this approach $m_{\rm pbc} \approx m_{\rm obc}$, a tremendous improvement.
However, that algorithm has a computational cost of $m^5$, making the net improvement
modest.

Here we introduce an improvement to this pbc MPS algorithm based on the
approximation of long products of certain large ($m^2 \times m^2$) transfer matrices
in terms of a singular value decomposition (SVD) with only a few singular values.
A new circular update procedure allows us to work exclusively with such long products.
Our approach improves the scaling of
the algorithm dramatically to $m^3$.

{\em MPS with pbc.} 
We summarize the algorithm presented in
Ref.~\onlinecite{verstraete:DMRG_and_pbc} and explain some practical aspects. 
The ground state of a quantum mechanical system like a spin model, defined on a one dimensional
lattice of $N$ sites, can  be written in 
terms of an MPS~\cite{garcia_MPS_review}
\begin{equation}
\label{eq:MPS}
  \ket{\phi} = \sum_{s_1,s_2\ldots s_N} \textrm{Tr}(\As{1}{s_1}\As{2}{s_2} \ldots \As{N}{s_N}) \ket{s_1 s_2 \ldots s_N},
\end{equation} 
where $\As{i}{s_i}$ are sets of $d$ matrices of dimension $m \times m$ and $d$ is the dimension of the
Hilbert space of a single spin $s_i$. 
The trace in eq.~(\ref{eq:MPS}) ensures periodic boundary
conditions. Any state can be written in this form if $m$ is large enough; the power of the
approach comes from the property that modest $m$ produces excellent approximations to ground states
of local Hamiltonians.
Of course the expression above is purely formal and we need a
procedure to optimize the matrices $\As{i}{s_i}$. 
For any operator $O_i$ on a site $i$ we define the $m^2\times m^2$ 
matrix ~\cite{ostlund:thermodynamic_limit_DMRG}
\begin{equation}
  \label{eq:map}
  E_{O_i}^{[i]} = \sum_{s,s'} \bra{s} O_i \ket{s'} \As{i}{s_i} \otimes \left(
    \As{i}{s_i'} \right)^*.
\end{equation}
Using these generalized transfer matrices, expectation values
of products of
operators can be easily evaluated
\begin{equation}
  \label{eq:prodE}
   \braket{\phi|\,O_1 O_2 \ldots O_N\,|\phi} = \textrm{Tr}(E_{O_1}^{[1]}E_{O_2}^{[2]}\ldots E_{O_N}^{[N]}).
\end{equation}
The Hamiltonian can also be written using the relation above and the matrices
$\As{i}{s_i}$ can be optimized one by one in order to minimize the energy. 
Consider the
Ising model \mbox{$H = \sum_i \sigma_i^z \otimes \sigma_{i+1}^z$}. To
optimize matrices $A_{s_i}^{[i]}$ at site $i$, an
effective Hamiltonian containing only matrices $\As{1}{} \ldots \As{i-1}{},\As{i+1}{}
\ldots \As{N}{}$ can be constructed as follows
\begin{equation}
  \label{eq:eff_ham}
  H_{eff} = \one^s \otimes \tilde{h}^i + \sigma^z \otimes \EprodT{i+1}{i-1}{l}
  +  \sigma^z \otimes \EprodT{i+1}{i-1}{r},
\end{equation}
where $\one^s$ is the identity matrix in spin space and
\begin{eqnarray}
\label{eq:E_products}
h^i & = & \sum_{k} E_{\simpleone}^{[i+1]} \ldots E_{\simpleone}^{[k-1]} E_{\sigma^z}^{[k]}
E_{\sigma^z}^{[k+1]} E_{\simpleone}^{[k+2]} \ldots E_{\simpleone}^{[i-1]}
\nonumber \\ 
\Eprod{i+1}{i-1}{l} & = & E_{\sigma^z}^{[i+1]} E_{\simpleone}^{[i+2]} E_{\simpleone}^{[i+3]} \ldots
E_{\simpleone}^{[i-1]} \\
\Eprod{i+1}{i-1}{r} & = & E_{\simpleone}^{[i+1]} E_{\simpleone}^{[i+2]} \ldots
E_{\simpleone}^{[i-2]} E_{\sigma^z}^{[i-1]}. \nonumber
\end{eqnarray}
In the equation above, all indices are taken modulo $N$. The tilde in
eq.~(\ref{eq:eff_ham}) refers to the exchange of indices $X_{(ij)(kl)} =
\tilde{X}_{(ik)(jl)}$.
Together with a map of the identity matrix
$N_{eff} \!=\! \one^s \otimes \tilde{N}^i$,
$N^i  \!=\!  E_{\simpleone}^{[i+1]} \ldots E_{\simpleone}^{[N]} E_{\simpleone}^{[1]} \ldots E_{\simpleone}^{[i-1]}$,
a new set of $d$ matrices $\As{i}{s_i}$ for fixed $i$
is found by solving the generalized
eigenvalue problem
\begin{equation}
  \label{eq:eigenvalue}
  H_{eff} \textrm{Vec}(A) = \epsilon N_{eff} \textrm{Vec}(A), 
\end{equation}
with $\epsilon$ the expectation value of the energy and $\textrm{Vec}(A)$
the $dm^2$ elements of $\As{i}{s_i}$, aligned to a vector. 

When a new set of matrices
has been found, the matrices need to be regauged, in order to keep the
algorithm stable. In DMRG this is not necessary since the basis of each
block is orthogonal. The orthogonality-constraint 
reads $\sum_{s_i} \As{i}{s_i} (\As{i}{s_i})^\dagger = \one $. It can be
satisfied as follows: The state is left unchanged when we substitute 
$\As{l}{s} \to \As{l}{s}X \equiv U^{[l],s}$ and 
$\As{l+1}{s} \to X^{-1}\As{l+1}{s}$, with some nonsingular matrix $X$. This
matrix $X$ has to be found such that $U^{[l]}_s$ obeys the normalization condition
$
  \sum_s U^{[l]}_s (U^{[l]}_s)^\dag = \one.
$
We obtain $X$
by calculating the inverse of the square root of $Q
=\sum_s \As{l}{s}(\As{l}{s})^\dag$.
Since
$Q$ is not guaranteed to be nonsingular, the pseudo-inverse has to be
used~\cite{verstraete:norm},
by discarding singular values close to zero in an SVD of $Q$.
$H_{eff}^i$ can be calculated iteratively~\cite{verstraete:DMRG_and_pbc}.
while updating the $A$-matrices one site at a time.
%
One sweeps back and forth in a DMRG like manner.

Vidal introduced a different approach, for infinitely long
translationally invariant
systems~\cite{vidal_translational_invariant}. By assuming only two different
kinds of matrices $A^{[1]}$ and $A^{[2]}$ and aligning them in alternating
order, an algorithm for both ground state and time evolution can be
constructed that updates the matrices in only $O(m^3)$ steps. However, unlike 
the periodic MPS method discussed here, Vidal's method does not apply to
non translationally invariant systems (e.g.\ when impurities or a
site dependent magnetic field are studied).  In addition, the periodic MPS method can be 
adapted \cite{porras:spectra_2005} to treat excited states, 
whereas the method of Ref.~\onlinecite{vidal_translational_invariant} probably cannot, since
the excitations would be spread over an infinite lattice and would have no effect on any individual site.
Recently, a related approach came to our attention \cite{Verstraete_m3},
in which the E-matrix of a translationally invariant system is treated in $O(m^3)$. 
Also recently, related Quantum Monte Carlo variational methods 
using tensor product states were introduced \cite{QMC_MPS1,QMC_MPS2}, 
with scaling $O(Nm^3)$ per Monte Carlo sweep.

{\em Computational Efficiency.}
It was shown in Ref.~\onlinecite{verstraete:DMRG_and_pbc} that the $m$ needed
for pbc systems in the MPS approach is comparable to the $m$ needed in
obc systems within DMRG.
However, 
it is also vital how CPU-time scales with $m$.
In efficient DMRG programs, most operations can be done by computing
multiplications of $m\times m$ matrices 
(see Ref.~\onlinecite{schollwock:DMRG_review}, Ch.~II.i).

In contrast, in the MPS-algorithm described above, operations on $m^2\times m^2$ matrices need to
be done to form the products of $E$-matrices that represent the
Hamiltonian. So one would expect the algorithm to be of order $O(m^6)$. 
By taking advantage of the special form of the $E$ matrices
eq.~(\ref{eq:map}), multiplications can be done in $O(m^5)$
which is, however, still $O(m^2)$ slower than DMRG. 

{\em Decomposition of products.}
We now introduce an approximation in the space of $m^2\times m^2$
matrices which reduces the CPU time dramatically while the accuracy of the
calculation does not suffer.
Let us perform a singular value decomposition of a long
product of $E$-matrices
\begin{equation}
  \label{eq:svd}
  E^{[1]}_{O_1}E^{[2]}_{O_2}\ldots E^{[l]}_{O_l} = \sum_{k=1}^{m^2} 
              \sigma_k \mathbf{u}_k^{\phantom{T}} \mathbf{v}_k^T.
\end{equation}

\begin{figure}[!tb]
  \centering
   \includegraphics[width=0.5\textwidth]{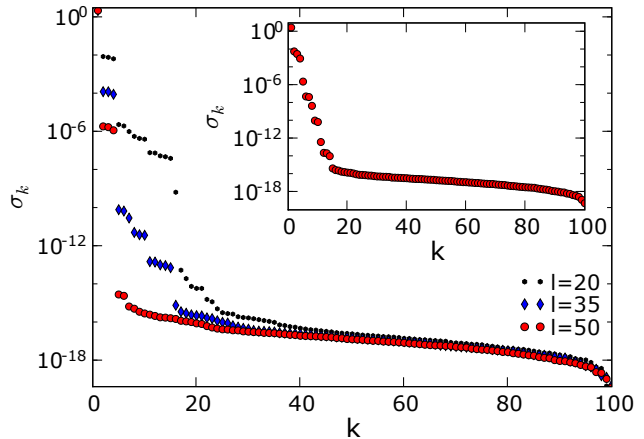}
  \caption{SVD of a product $E^{[1]}_{\simpleone} \dots
    E^{[l]}_{\simpleone}$ with $m = 10$. The logarithm of the singular
    values $\sigma_k$ is shown for different $l$ in the case of a spin $1$
    Heisenberg chain of length $100$ with periodic boundary conditions. The
    inset shows data for a spin $\frac{1}{2}$ Heisenberg chain.
    }
  \label{fig:svd}
\end{figure}

It turns out that the singular values $\sigma_k$ decay very fast. This is
shown in Fig.~\ref{fig:svd} for products of the form $\prod_{i=1}^l
E_{\simpleone}$ 
with various values of $l$,
for the case of the spin $1$ Heisenberg chain.
One can see that the longer the product
the faster the singular values decay,
roughly exponentially in the length $l$.
We therefore propose to approximate long products in a reduced basis
\begin{equation}
  \label{eq:svd_approx}
  \prod_{i=1}^l E_{O_i}^i \approx \sum_{k=1}^{p} \sigma_k
  \mathbf{u}_k^{\phantom{T}} \mathbf{v}_k^T,
\end{equation}
with $p$ chosen suitably large. In the example of Fig.~\ref{fig:svd}, we
would choose $p$ to be $4$ at $l=50$. Remarkably, for longer products $p$
can be as small as $2$ without a detectable loss of accuracy. Thus, the large
distance behaviour of the ground state of the spin $1$ chain is encoded in
these two numbers, similar to the transfer matrices of a classical spin
chain.
The situation does not change significantly when more complicated operators
such as the Hamiltonian are decomposed. Of course, the decay of the singular
values will be model dependent. For a spin $\frac{1}{2}$ Heisenberg chain 
we found that the decomposition can be done in the same manner with 
approximately the same number of singular values to be kept.

A multiplication of a product with a new $E$ matrix
can therefore be done%
\footnote{Denote $\mathbf{v}$ as an $m^2$ vector and $V$
as the  elements of $\mathbf{v}$ aligned as an $m\times m$ matrix. Then one
can use the relation $(A \otimes B) \mathbf{v} = Vec(B V A^T)$ to perform a
matrix-vector product.} 
in $O(p m^3)$ and a multiplication 
of two terms like~(\ref{eq:svd_approx}) can be done in
$O(p p' m^2)$. By building the effective Hamiltonian out of products in
this representation, the
iterative evaluation of the eigenvalue problem can be accelerated. Whereas in a
dense form each matrix-vector multiplication -- which occurs in eigenvalue routines such
as Lanczos or Davidson -- takes $(d m^2)^2$ operations, it can now be done in
$O(d^2 p m^2)$. 
Note that all operations are now done on matrices of size \mbox{$m \times m$}.

{\em Performing the SVD in $m^3$.}
A crucial step is the efficient generation of the SVD representation of a large 
$m^2\times m^2$ matrix $M$ in only $O(m^3)$ operations.
We describe a simple algorithm, with 
a fixed number of singular values (four) to keep the notation simple.
Suppose that 
$M = U d V$, with $d$ a $4\times4$
diagonal matrix, and
that multiplication of $M$ by a vector (without using the SVD factorization)
can be done in $O(m^3)$.
To construct $U$, $d$, and $V$ 
with $O(m^3)$  operations,
we first form a random $4\times m^2$ matrix $x$,
and construct $y = x M$. The $4$ rows of $y$,
are linear combinations of
the rows of $V$.  Orthonormalize them to form $y'$.
Its rows 
act as a complete orthonormal basis for the rows of $V$.  This means that
$V = V y'^T y'$, and thus $M = M y'^T y'$. Construct $z = M y'^T$, and perform
an SVD on $z$:  $z = U d V'$.  Then $M = z y' = U d V$, where $V = V' y'$. $V$ is
row orthogonal because $V'$ is orthogonal and $y'$ is row orthogonal.
The calculation time for the orthogonalization of $y$ and the SVD of $z$ is $O(m^2)$,
and so the calculation time is dominated by the two multiplications by $M$, e.g.\
roughly $2 \times 4 \times O(m^3)$.%
\footnote{A Lanczos approach would take an additional factor for convergence.}

In applying this approach to the periodic MPS algorithm, $M$ is a product of $O(N)$ 
$E$-matrices like in  eq.~\ref{eq:E_products}, which in turn are outer products (\ref{eq:map}).
The multiplication with $M$ can be done iteratively in $O(N m^3)$ operations,
analoguously to the construction of $H^i_{eff}$.
The calculation time is thus $O(N m^3)$ for each SVD representation 
generated this way.  It is only  needed a few times per sweep (see below).

{\em A circular algorithm.}
A speed-up in the simulation can only be expected if the number of singular
values that need to be included is sufficiently small. 

However, in the algorithm of Ref.~\onlinecite{verstraete:DMRG_and_pbc} one
sweeps back and forth through the lattice, so that close to the turning points,
products of only a few E-matrices appear,
which require more singular values 
(Fig.~\ref{fig:svd}).
In the extreme case of only one $E$-matrix, we would have $p = m^2$. 
%
%
To overcome this bottleneck 
we propose a modified method which proceeds through the chain in a
{\em circular} fashion,
thus making natural use of the periodic boundary conditions.
Note that we cannot employ multiplications with inverse matrices $E_O^{-1}$,
since they are too expensive to calculate.
We consider the lattice as a circular ring, and divide it into thirds,
or ''sections".  We perform update steps
for one section at a time.  
To start one section, we first construct the
Hamiltonian and other necessary operators (see eq.~\ref{eq:E_products})
corresponding to the other two sections of the lattice.
Only a few such operators are needed.
Each of them contains products of $N/3$ $E$-matrices 
and is computed by an SVD decomposition as described 
before.

Then a set of these operators is made by successively adding sites from
the right most part of the current section
to the operators constructed for the section on the right,
working one's way to the left.  
Adding a site involves the multiplication of an E-matrix to the left of an operator.
These steps can each be done in $O(m^3)$ operations.
When one has reached the left side of the current section,
its initialization is finished and
one can start the normal update steps, now building up a set of
operators from the left,
again in $O(m^3)$ operations. 
One stops when one reaches the right hand side of the current section.
Then the procedure repeats with the section to the right as the new current section.
Some of the operators previously computed can be reused.
The updates now go in a circular
pattern rather than the usual back and forth.

By proceeding in this way on a system of length $N$, the
blocks on which we have to perform an SVD are of length at least $N/3$ (if we
split our system into three parts), so that 
the SVD will have only few singlular values.
Consequently,
the algorithm is expected to scale like $O(N m^3)$.

\begin{figure}[!tb]
  \centering
     \includegraphics[width=0.52\textwidth]{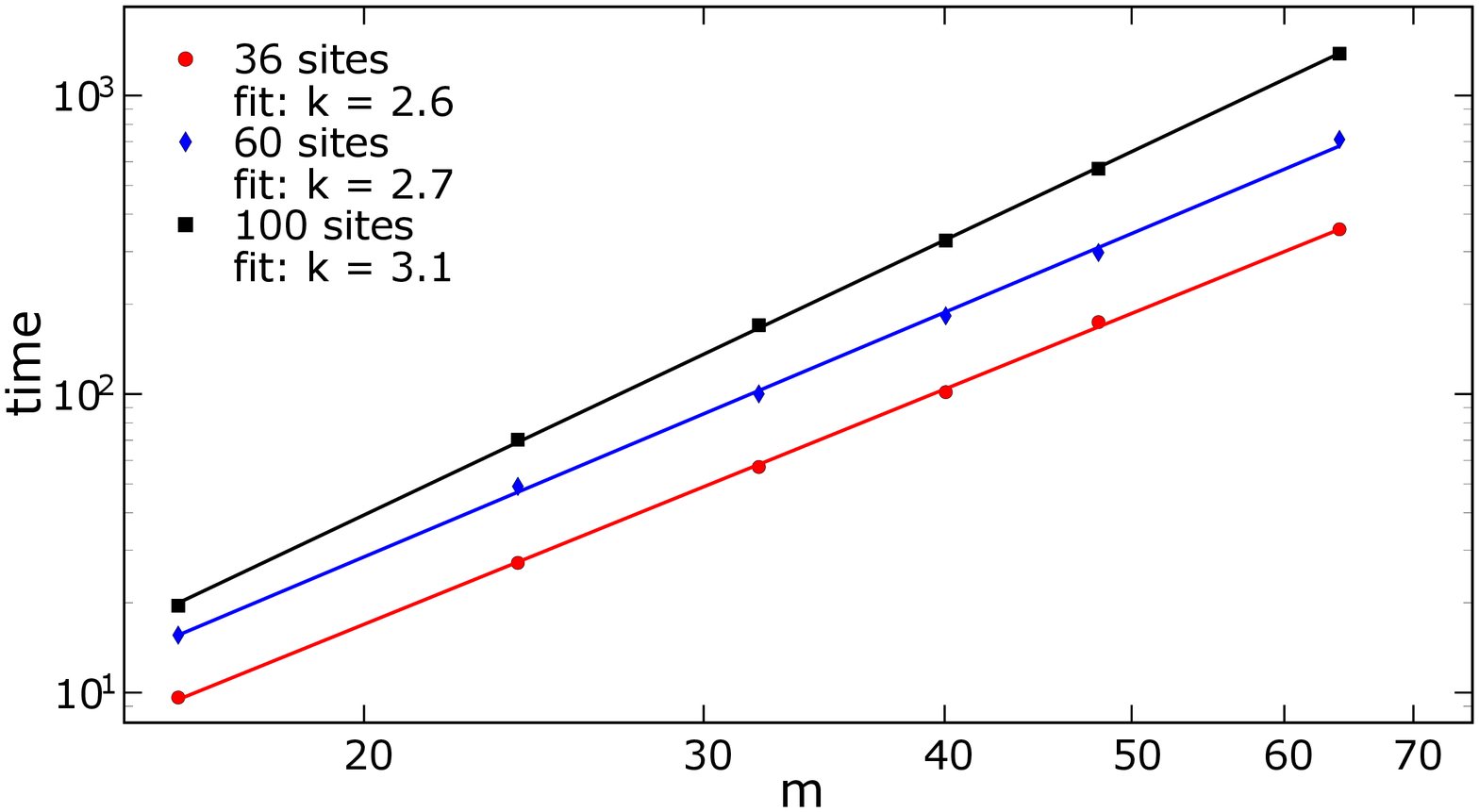}
    \caption{Scaling of the circular version of the algorithm. CPU--time per
      sweep (one update of each site) is measured on a 100 site spin $\frac{1}{2}$ Heisenberg chain
      for different system sizes. The time is fitted to a function $m^k$.}
    \label{fig:time}
\end{figure}

{\em Test and Results.}
To test our improvements, we studied spin $1$ and spin $\frac{1}{2}$ Heisenberg
chains up to length $N=100$.

The exact ground state energy 
for pbc on an \mbox{$N=100$} chain in the spin $1$
case is found to be \mbox{$E_0/N \cong -1.4014840386(5)$} via a DMRG calculation with
$m = 2000$. 
The error is generously estimated from the truncation error and an extrapolation in $m$.
The periodic result differs from the infinite system result (determined using long open chains)
only in the last decimal 
place, so we will call  this value ``exact''.

We discarded singular values smaller than a $10^{-11}$th of the 
largest one. This parameter is chosen such that the algorithm remains
stable, which is not the case if the error bound is chosen too large
($10^{-8}$ or larger).
To decrease the time it takes until convergence is reached, we start our
calculation with small $m$ and enlarge it after the algorithm converged for
the current $m$. This is also done in many DMRG programs.
We enlarge the matrices $A$ and increase their rank by filling the
new rows and columns with small random numbers $r$, uniformly
distributed in the
interval $[-10^{-6},10^{-6}]$. The
number of sweeps it takes until convergence is reached is similar to DMRG. 
For the present model, two or three sweeps are enough for each
value of $m$.

Fig.~\ref{fig:time} shows that the algorithm indeed scales like $m^3$,
and no longer like $m^5$.
It is slightly faster on small systems, due to faster parts of the algorithm,
and becomes slightly slower on large systems, likely due to memory access times.
Our method (on a periodic system) requires a constant factor of about 10 
as many operations per iteration as DMRG does on an open system,
which is still very efficient.

\begin{figure}[!tb]
  \centering
  \includegraphics[width=0.5\textwidth]{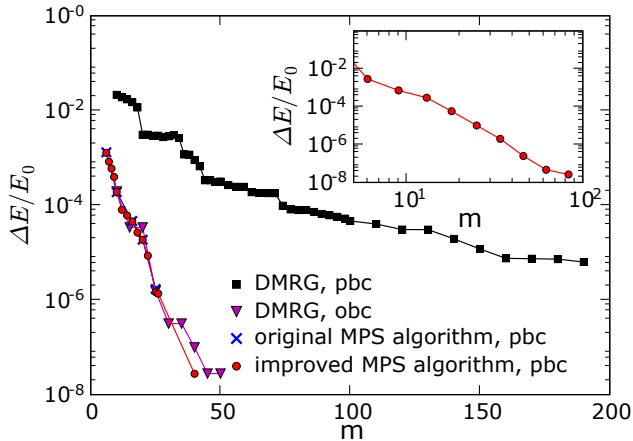}
  \caption{Relative error of the ground state energy of the spin $1$
    Heisenberg model versus the dimension $m$ of the reduced Hilbert space. 
    DMRG results with obc and pbc
    are shown, as well as Matrix Product State results with pbc.
    The inset shows MPS results with pbc for a spin $\frac{1}{2}$ Heisenberg chain of 100 sites.
    }
  \label{fig:gs_energy}
\end{figure}
Finally, we studied the convergence to the exact ground state energy as a function of $m$. 
We investigated DMRG with obc and pbc, and the MPS algorithm with pbc, both the
original version and our improved method. 
The relative error $\frac{\Delta E}{|E_0|}$  
for these cases is plotted in Fig.~\ref{fig:gs_energy}.
The relative error of the spin correlation function (not shown)
is of similar magnitude with our improved method. 

As has been well known, DMRG with obc performs much better than with pbc. 
With the MPS algorithm and pbc the relative error as a function of $m$ is comparable to the
error made with DMRG and obc. This has already been reported earlier~\cite{verstraete:DMRG_and_pbc}.
The important point here is that the error remains the same when we introduce the
approximations. 
Also, the number of sweeps until convergence is reached is similar for DMRG with obc and for MPS.
We note that the convergence in Fig.~\ref{fig:gs_energy} is consistent with exponential behavior in the spin $1$ case
and with a power law for spin $\frac{1}{2}$.

In a typical DMRG calculation, matrix dimensions $m \simeq 100 - 1000$ (and larger) are used.
To illustrate the computational time scaling, suppose we study a model which requires $m=300$
states for obc with traditional DMRG.
Then our new approach gains a factor of 
roughly $m^5/m^3\simeq 10^5$ over the method of Ref.~\onlinecite{verstraete:DMRG_and_pbc},
and even more over traditional DMRG.

%
In summary, by introducing a well controlled approximate representation
of products of MPS transfer matrices in terms of a singular value decomposition,
we have formulated a circular MPS method for systems with
periodic boundary conditions, which works with a computational
effort comparable to that of DMRG with open boundary conditions.

\acknowledgments
We acknowledge  support from the NSF under grant DMR-0605444 (SRW)
and from NAWI Graz (PP).


\bibliographystyle{h-physrev}

\end{document}